\documentclass[twocolumn,floatfix]{openjournal}

\usepackage{lipsum}

\usepackage{xcolor}
\usepackage{textgreek}
\usepackage[utf8]{inputenc}
\usepackage[english]{babel}

\usepackage{hyperref}
\hypersetup{
    unicode, 
    colorlinks=true,
    linkcolor=linkcolor,
    citecolor=linkcolor,
    filecolor=linkcolor,
    urlcolor=linkcolor,
}
\usepackage{color,colortbl}
\definecolor{linkcolor}{rgb}{0.0,0.3,0.5}
\usepackage{tensind}
\tensordelimiter{?}
\DeclareGraphicsExtensions{.bmp,.png,.jpg,.pdf}
\usepackage{verbatim}
\usepackage[normalem]{ulem}
\usepackage{orcidlink}
\usepackage{soul}

\graphicspath{ {./figs/} }

\newcommand{\msun}{{\rm M}_\odot}

\usepackage{orcidlink}
\usepackage{autobreak}

\begin{document}

\title{Feedback shaped the galaxy morphological sequence in presence of mergers}

\author{Masafumi Noguchi\,\orcidlink{0000-0003-3585-4631}$^{1}$}

\affiliation{$^1$Astronomical Institute, Tohoku University, Aoba-ku, Sendai 980-8578, Japan}

\email{noguchi@astr.tohoku.ac.jp}

\begin{abstract}
Bulges and disks are major structural components that define galaxy morphology.
 The mass ratios of bulges and disks increase statistically with the galaxy mass, with the high-mass end occupied by elliptical galaxies. 
Although previous theoretical studies have succeeded in reproducing this morphological sequence, 
it is not yet fully understood why and how this morphological sequence emerged.
 Galaxy mergers accompanying dark matter halo mergers have been proposed as the major route for bulge formation. 
On the other hand, it is observationally known that 
 the mass fraction of galaxies (stars plus cold gas) in dark matter halos attains 
the peak value at 
 $M_{\rm halo} \sim 10^{12} {\rm M}_\odot$ throughout the cosmic time.
Using a simple galaxy evolution model including mergers, we show that this feature is the fundamental cause of the morphological sequence.
Halos hosting massive galaxies, which stay more massive than this peak mass for long periods during their growth,  
 merge mostly with satellite halos having larger galaxy mass fractions than themselves.
 Such mergers increase the bulge mass fraction efficiently. 
In contrast, host halos of low-mass galaxies evolve under unfavorable condition for 
bulge growth because they stay below the peak mass and merge with satellite halos 
with smaller galaxy mass fractions. 
Previous studies suggest that the peak in galaxy mass fraction is created by
feedback processes from active galactic nuclei (AGN) and young massive stars including supernovae (SN), which are considered to suppress star formation in high-mass and low-mass galaxies, respectively. 
This study thus points to a close relationship between the galaxy morphology and feedback processes which have hitherto been considered unrelated and suggests the importance of further investigation into their causal relationship.
\end{abstract}

\section{Introduction}

Galaxies exhibit a large diversity in their morphology, which is often characterized by the mass ratio of central spheroidal bulges and surrounding flat disks. 
As the galaxy mass increases, the mass fraction of bulge generally increases, 
up to unity for elliptical galaxies \citep{fisher2011,2011ApJS..196...11S,2014ApJS..210....3M}. 
This correlation, known as the galaxy morphological sequence, is one of the most notable relationships between various properties of galaxies, dating back to the historical 
Hubble sequence. 
 Cosmological simulations \citep[e.g][]{vogelsberger2013,Nelson2015,pillepich2018}
 and semi-analytic models \citep[e.g.][]{Cole2000,Cattaneo2006,lacey2016,Lagos2018} 
have attained considerable success in reproducing the morphological sequence 
in addition to the correlation between other properties such as galaxy stellar masses, cold gas contents, 
dynamical and kinematical structures, ages, metal contents of disks and bulges.
This success opened and promoted  galactic archaeology, which aims to get  new insight into the origin of 
various galactic components based on the observed properties of individual stars in galaxies 
\citep[see a recent review in ][]{deason2024}.

The most promising mechanism for bulge formation is mergers of galaxies.
At the most basic level, past simulations of two merging disk galaxies  indicate that the remnants have similar internal mass distributions to 
the observed profiles of bulges and elliptical galaxies \citep{1988ApJ...331..699B,Naab1999,Hopkins2009,2015MNRAS.452.4347K}. 
A number of extensive studies that treat evolution of galaxy populations as a whole suggest 
 that the galactic bulges have been formed by mergers of smaller galaxies 
into more massive galaxies \citep{Oesch2010,Avila-Reese2014,2008ApJ...672..177L}. 
Despite past success in reproducing many observational properties of galaxies, the fundamental mechanism for the emergence of the 
morphological sequence remains unclear.

On the other hand,
it is well established that the stellar-to-halo mass ratio (SHMR) has a peak at the halo 
mass around $10^{12} {\rm M}_\odot$ through most of the cosmic time \citep[e.g.][]{Behroozi2013,rodriguez2015,moster2018}.
Not the stellar mass but the galaxy mass including cold interstellar medium obeys 
a similar relation,
namely, the galaxy-to-halo mass ratio (GHMR) 
has a peak that stays at $M_{\rm halo} \sim 10^{12} {\rm M}_\odot$ through the time
 \citep[e.g.][]{Popping2015,guo2023}.
Here, using a simple model of galaxy evolution including mergers, we argue that this feature of GHMR plays a key role in shaping the galaxy morphological sequence through mergers.
We employ here GHMR instead of SHMR.
This ratio, including cold gas contents in galaxies in addition to stars, represents 
the dominance of total baryonic contents in halos. We prefer this option because cold gas 
also contributes to the bulge formation by inducing starbursts in mergers.

Past studies suggest that the feedback processes by young massive stars including supernovae and active galactic nuclei 
(AGNs) are promising candidates to create the peak in SHMR and GHMR.
The former is considered to be effective in suppressing star formation in low-mass galaxies by expelling or heating ISM
 \citep[e.g][]{Dekel1986,maclow1999,Scannapieco2008,stinson2013}.
The latter is expected to operate preferentially in massive galaxies, in which the radiation and jets from supermassive black holes can heat the halo gas to high temperatures, 
 hindering or delaying  the accumulation of ISM  
 \citep[e.g][]{bower2006,croton2006,somerville2008,fabian2012,moster2013,vogelsberger2013,hopkins2014,torrey2014}.
As a logical consequence, we propose here that the morphological sequence has a causal relationship with 
feedback processes.

We describe our evolution model in section 2. Section 3 compares our models 
 with the observational data regarding the bulge-to-total 
stellar mass ratios. Section 4  examines how the GHMR controls the bulge growth 
depending on  the galaxy mass. 
Bulges grow through multiple pathways in mergers. Section 5 examines the formation and 
assembly history of bulge stars for each pathway, followed by conclusions in section 6.

\section{Method}

Our method, like the semi-analytic models (SAM) \citep[e.g.][]{lacey2016,croton2016,mitchell2018} and the equilibruim models \citep[e.g.][]{dave2012,mitra2015},  does not rely on numerical 
simulations but traces galaxy evolution analytically except introducing simulation results 
as input physics. Unlike SAM, we do not address the populational properties such as galaxy mass functions, luminosity functions, and spatial correlation functions.
We instead follow the galaxy evolution for various halo masses specified at the present epoch by time integration 
from the appropriate initial condition.
In addition to the mass growth of dark matter halos, we analytically model fundamental baryonic processes 
such as accretion of halo gas to forming galaxies, star formation from accreted interstellar medium (ISM), ejection of ISM  by feedback processes,
and recycling of the ejected ISM. 
Effect of AGN is handled phenomenologically by modifying the accretion rate of the halo gas.
We permit non-equilibrium states of the system unlike the equilibrium approach.
For example, the mass of the interstellar medium is determined by the balance of the replenishment by accretion from halos 
and the consumption by star formation and therefore varies with time generally.
Our modelling is admittedly idealized, missing many details included in cosmological simulations and SAM. 
This approach, however, helps grasp the effect of each physical process
without delving into 
complexities of galaxy evolution. 

Modelling in this work comprises two steps. First, we calculate the evolution of isolated states in which
galaxies evolve under gas accretion from the intergalactic space with no mergers included.
In other words, we assume that the growth of  halos proceeds only through smooth accretion 
of diffuse dark matter followed by gas accumulation into growing dark matter halos.  
This step is performed to provide the first order approximation for galaxy properties  
required as input in the second step described below, which includes halo mergers.
The quantities required  include the masses of dark matter, hot halo gas, stars,  cold interstellar gas in
the galaxy, and the star formation history (SFH) of the stellar component (i.e., the age distribution of constituent stars).

\subsection{Calculation of isolated galaxy evolution}

Detailed description of the first step is given in a previous series of papers \citep{Noguchi2020,Noguchi2021,Noguchi2022,Noguchi2023}, 
which investigate the evolution of isolated galaxies in various accretion schemes of the halo gas.
We here repeat essential ingredients of this step.
The starting point is the time evolution of the halo mass for a given value of the present mass.
We use the parametrization of the mass growth history given in APPENDIX H of \citet{Behroozi2013},
 which fits very well with the results of Bolshoi \citep{klypin2011}, MultiDark \citep{klypin2016}, and Consuelo  
cosmological dark matter simulations.
Once the halo growth history is determined, we calculate the mass accretion rate of gas into halos 
by assuming the universal baryon fraction of $f_{\rm b} =0.17$.
We assume that when the mass of a halo increases by $\Delta M_{\rm halo}$, its portion $f_{\rm b} \Delta M_{\rm halo}$ is attributed to the increment of the halo gas.

We next calculate the accretion rate of the halo gas onto the galaxy on the basis of cold accretion paradigm 
by \citet{dekel2006}. This paradigm summarizes the results of hydrodynamical cosmological simulations 
including cooling and heating processes of gaseous components and describes how the behavior of the halo gas
depends on the halo mass and redshift \citep[e.g.][]{keres2005,Ocvirk2008,vandevoort2011}.
In model calculation, the freefall time and the radiative cooling time of the halo gas are
calculated at each timestep. The former is simply determined by the mean mass density of the halo, 
 which is in turn determined by the cosmological mean density 
for the specified cosmology. The latter is calculated from the density, temperature 
 and metallicity of the halo gas,  using the table for 
the radiative cooling time under the collisional excitation equilibrium given in  
\citet{sutherland1993}.
The halo gas is assumed to accrete to the galaxy and becomes a part of the cold gas (ISM) with the free-fall time
when the halo mass satisfies the condition $M_{\rm halo} \leq M_{\rm shock}$.
Here, $M_{\rm shock}$ is the mass scale below which the halo gas is unable to develop a stable shock wave and rapid radiative cooling induces a free-fall accretion of the halo gas \citep{Birnboim2003,Whitaker2021}.
When the halo mass satisfies the condition
 $M_{\rm shock} < M_{\rm halo} \leq M_{\rm stream}$, 
half the halo gas is assumed to accrete with freefall time and another half is assumed to
accrete with the radiative cooling time.
$M_{\rm stream}$  is the mass scale above which the shock-heating and AGN feedback are expected to destroy filamentary accretion of cold halo gas to the center \citep{dekel2006,Ocvirk2008}.
When the above condition is satisfied, the halo has a composite structure comprising cold 
unheated gas and the shock-heated hot gas. The recipe taken here is meant to mimic
this situation in a simplified manner.
When the halo mass satisfies the condition $M_{\rm halo} \geq {\rm max}(M_{\rm shock},M_{\rm stream})$, 
the halo gas is assumed to be totally heated by shock waves to the virial temperature and 
accrete with the radiative cooling time.
Here we take $M_{\rm shock}= 10^{11.7} {\rm M}_\odot$ and $M_{\rm stream} = 10^{11.25+0.42z}  {\rm M}_\odot$,
where the latter is valid only if $M_{\rm shock} \leq M_{\rm stream}$
 and $z$ denotes redshift.
These mass scales were chosen on the basis of results of the above cosmological simulations and previous semi-analytic studies that explored the role of AGN feedback in star formation quenching \citep{Cattaneo2006,croton2006}.

We briefly describe the treatment of baryonic processes associated with the accreted gas.
See \citet{Noguchi2023} for more details.
We follow \citet{blitz2006} in implementing star formation process.
The gas accreted from the halo is assumed to form a gaseous disc and replenish the interstellar medium 
(ISM), from which stars are formed. 
Star formation rate density, $\Sigma_{\rm SFR}$, is given by
\begin{equation}
	 {\frac { \Sigma_{\rm SFR} }  {(\rm M_{\odot} pc^{-2} Gyr^{-1})} } = 
	 {\frac {\epsilon_{\rm SF}}  {(\rm  Gyr^{-1})}}  { \frac {\Sigma_{\rm HCN}}
	   {(\rm M_{\odot} pc^{-2})}}  
\end{equation}
where $\epsilon_{\rm SF} = 13 {\rm Gyr^{-1}}$
  \citep{gao2004,wu2005} and $\Sigma_{\rm HCN}  $
stands for the surface density of the dense molecular gas.
Massive stars provide supernova feedback to the surrounding ISM  
and eject a part of it.
The ejected mass  is calculated as
\begin{equation}
	 \Delta M_{\rm ej} = \left[{ \frac {2 \epsilon_{\rm FB} E_{\rm SN}  \eta_{\rm SN}}
	   {V_e^2}} \right] \Delta M_{\rm star}
\end{equation}
where $\Delta M_{\rm star}$ is the mass of stars formed in the current time step,
$\epsilon_{\rm FB}$ is the feedback efficiency,
$E_{\rm SN}$ is the energy released by one supernova, and $\eta_{\rm SM}$ is the
number of supernovae produced per  one solar mass of stars formed at that step.
We take $E_{\rm SN}=10^{51}$ erg and $\eta_{\rm SN}=8.3 \times 10^{-3}$ following \citet{dutton2009}.
For feedback efficiency we adopt $\epsilon_{\rm FB}=0.2$ as the fiducial one.
One caveat here is that we do not include radiation pressure from massive stars.
It could contribute significantly to feedback process, especially reducing star formation rates in low-mass galaxies by
heating ISM \citep[e.g.][]{andrews2011,rosdahl2015,hopkins2020}.
It is a future task to include this effect in star formation recipe in a consistent way.

The ultimate fate of the ejected gas is highly uncertain. We assume that some portion 
of the ejected gas falls back and returns to the galactic disc for further star formation
\citep[e.g.][]{oppenheimer2010,henriques2013,christensen2016}.
We use the parametrization by \citet{mitra2015} for the timescale of reincorporation (recycling) 
of the ejected gas as
\begin{equation}
	t_{\rm rec}  = 0.52 \times 10^{9} {\rm yr} \times (1+z)^{-0.32}
	\left( \frac {M_{\rm vir}}  {10^{12}{\rm M}_{\odot}}  \right)   ^{-0.45}
\end{equation}
This is obtained as the best fit for their equilibrium model. 
The mass recycled at each time step is calculated as 
\begin{equation}
	M_{\rm rec}  = \left( M_{\rm ej} / t_{\rm rec}\right)  \Delta t
\end{equation}
where $M_{\rm ej}$ is the total mass of ejected material present at the current time step
(i.e., the reservoir of the ejected material) 
and $\Delta t$ is the (constant) length of a single time step.
It should be noted that $M_{\rm ej}$ increases by the stellar ejective feedback 
and decreases by recycling. Therefore, it is updated at each time step as
\begin{equation}
	M_{\rm ej} = M_{\rm ej} + \Delta M_{\rm ej} - M_{\rm rec}
\end{equation}

In low-mass galaxies, the gas accretion is assumed to be suppressed by the preventive feedback accompanying
galactic winds driven by the supernova feedback.
This type of feedback was introduced to solve a tension between mass and metallicity 
in local low-mass galaxies in semi-analytic models \citep[e.g.][]{lu2017}.
Cosmological simulations show that outflowing winds indeed cause the preventive 
feedback
 \citep{vandevoort2011,Mitchell2020,mitchell2022}.
To mimic the preventive feedback in a simple form, 
we reduce the halo gas accretion rate to the disc calculated as above to  the level
$$ f_{\rm p} = f_{\rm pr}+(1-f_{\rm pr})(y+\frac{\pi}{2}) $$,
\noindent{where }
$$ y={\rm arctan}[3({\rm log} M_{\rm vir}-{\rm log} M_{\rm pr})/w_{\rm pr}] $$
\noindent{Here} $f_{\rm pr}$ is the parameter controlling the strength of the feedback, $M_{\rm pr}$
is the critical mass below which the feedback is significant, and $w_{\rm pr}$ controls
how suddenly the feedback gets strong with the decreasing halo mass. 
$M_{\rm pr}$ is a function of redshift and given by
$${\rm log} M_{\rm pr}=M_0+z(M_5-M_0)/5, $$
\noindent{where} $M_0$ and $M_5$ are the critical mass at $z=0$ and 5, respectively.
We take $f_{\rm pr}$=0.3, $M_0$=11.8, $M_5$=11.0, and $w_{\rm pr}$=0.5 as the default values.
For this choice, $f_{\rm p} \sim 0.65$ at the critical mass.
The original accretion rate is multiplied by the calculated factor $f_{\rm p}$ to model the preventive feedback.
See \citet{Noguchi2023} for details about choice of these parameters.

\subsection{Calculation including mergers}

The second step introduces mergers of dark matter halos. The more massive and less massive halos in a merger are called the primary and satellite, respectively. 
The probability with which a particular primary merges with a satellite having a given mass at a given time is calculated based on the empirical data obtained from the cosmological N-body simulations by \citet{Fakhouri2010}.
This work gives the merger rate {\it per descendent halo}.
We need the rate {\it per progenitor halo} to trace the mass growth of halos by 
forward time integration.
We convert the former into the latter using the method given in APPENDIX of \citet{genel2009}.
 Halos also grow by diffuse accretion of dark matter from outside.
To calculate the contribution of this accretion, we first specify the growth of the total mass of a halo based on the result from
cosmological numerical simulations as done in the first step and then subtract from the total growth rate
the portion contributed by mergers. We set the lower limit of $\xi=0.03$ for the satellite-to-primary halo mass ratio.

At each time step, we first pick up mergers at random in accordance with the  merger probability per progenitor calculated as above.
 If a particular merger is picked up, we  calculate the merger time \citep{genel2009}, i.e., the time needed for two halos to merge into a single halo, as

\begin{equation}
T_{\rm merger} =0.7 \xi^{1.3/(ln(1+\xi))} H({\it z}=0)/H({\it z}) {\rm Gyr},
\end{equation}

\noindent{where $H$ denotes the Hubble parameter. If the merger is judged not to be completed by the present cosmological epoch, it is discarded.}
In addition to the halo mass growth due to mergers, we also add the contribution from the
diffuse accretion mode calculated as explained above at every time step.

We assume that galactic disks form by star formation in ISM and bulges form only through mergers.  To calculate increment of the bulge mass in a particular merger, 
we need to know the properties of the satellite galaxy and the primary galaxy  
and to apply an appropriate recipe to calculate the properties of the merger remnant.
The merger recipe we adopt is described below.
In the first iteration, we use the evolution data of galaxies obtained in the isolated run described above as the input for the satellite properties.
The first iteration run records the physical quantities of all the galaxies in the same manner as the isolated run, preparing the input data for the second iteration.
In this way, the $N$-th iteration is performed using galaxy properties obtained in the ($N-1$)-th run until the convergence is reached.

\begin{figure*}
        \centering
        \includegraphics[width=1\linewidth]{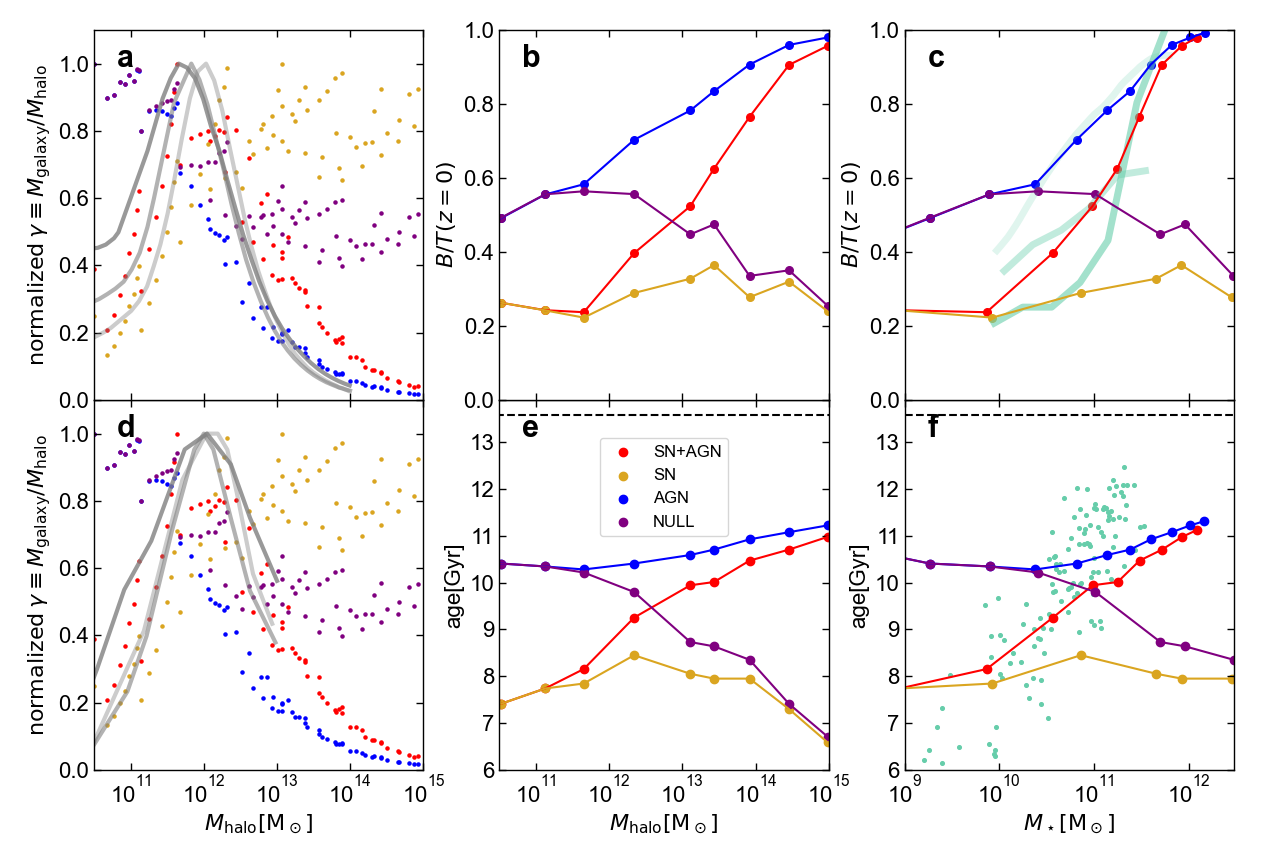}
        \caption{Effects of feedback and galaxy-to-halo mass ratios on the bulge mass fraction and age. Models SN+AGN, SN, AGN, and NULL are colored by red, brown, blue and purple, respectively.
{\bf a} and {\bf d}: The mass ratio, $\gamma$, of the galaxy and its halo is plotted against the halo mass.
Gray lines indicate observations by \citep{Popping2015} ({\bf a}) and \citep{guo2023} ({\bf d}) for $z=$2 (dark), 1 (medium), and 0 (light), respectively.
Each plot is normalized so that its maximum value is unity.
{\bf b} and {\bf c}: Mass ratio, $B/T$, of bulges and total stellar content of galaxies at present are plotted against the halo mass ({\bf b}) and the galaxy stellar mass ({\bf c}).
Green lines in {\bf c} denote the data for MaNGA galaxies \citep{Bundy2015} (dark), SDSS galaxies sampled by \citet{2011ApJS..196...11S}(medium),
and SDSS galaxies sampled by \citet{2014ApJS..210....3M} (light), respectively. {\bf e} and {\bf f}: Same as {\bf b} and {\bf c}, respectively, but for ages of bulges.
Here, age is the lookback time at which 50$\%$ of stars were formed. Green dots in {\bf f} indicate the observation for local galaxies by \citet{Breda2018}.
        }
        \label{fig1}
\end{figure*}

\subsection{Galaxy mergers}

Bulges generally grow in three ways through mergers \citep[e.g.][]{delucia2007,Hopkins2009}.
First, they acquire stars from satellite galaxies.
Second, the stellar disk of the primary is partly destroyed and absorbed into the bulge by violently changing gravitational fields in mergers.
Third, mergers induce active star formation (starburst) from cold interstellar gas of both galaxies and stars formed are added to the bulge of the primary galaxy.
These processes are taken into account in many semi-analytical studies for galaxy formation \citep[e.g.][]{somerville1999,croton2006,Lagos2018}.
 The contribution of each pathway is determined by the properties of galaxies contained in the primary and satellite halos.
We follow the simulation results for two merging galaxies by \citet{Hopkins2009} in computing the amount of stars formed in each pathway.

The stars in the satellite galaxy are all added to the bulge of the primary galaxy.
The age records of satellite stars (namely SFH of the satellite galaxy) are also 
added to the SFH of the primary bulge.
A portion of those stars may actually settle in the stellar halo of the primary, possibly leading to overestimate
of bulge masses in our models. Past observational and theoretical studies, however, suggest that the mass fraction of stellar halos
is under $\sim 20$ percent of total stellar masses \citep[e.g.][]{Behroozi2013,elias2018,behroozi2019} and remains subdominant in the total stellar budget.
We think therefore that neglect of stellar halos  is acceptable in the present level of model calculation.

Second, the part of the primary stellar disk with the same mass as the added satellite stars is converted to the bulge, and its mass is subtracted from the stellar disk
(If mass of the added stars exceeds the primary stellar disk mass, the latter is totally added to the bulge).
The age record of the converted disk (which is assumed to be the same as that of the whole disk up to that time) is added to the bulge age record.

Third, the stellar mass formed in merger-driven bursts is calculated as follows.
Hereafter, the stellar and gaseous masses of a galaxy are denoted by 
 $M_\star$ and $M_{\rm g}$, respectively.
Subscripts '$_{\rm p}$' and '$_{\rm s}$' mean the primary and satellite quantities, respectively.
The burst stellar mass is calculated as

\begin{equation}
M_{\star,{\rm burst}} =2(1-f_{\rm gas})  \frac{\mu}{1+\mu}(M_{\rm g,p}+M_{\rm g,s})
\end{equation}

\noindent{where $f_{\rm gas}$ is the gas fraction defined by }

\begin{equation}
f_{\rm gas} \equiv  (M_{\rm g,p}+M_{\rm g,s})/(M_{\star,{\rm p}}+M_{\star,{\rm s}}+M_{\rm g,p}+M_{\rm g,s})
\end{equation}

\noindent{
Here, the galaxy mass ratio is defined as
}
\begin{equation}
\mu \equiv (M_{\star,{\rm s}}+M_{\rm g,s})/(M_{\star,{\rm p}}+M_{\rm g,p})
\end{equation}
\noindent{This burst mass is added to the bulge of the primary galaxy whereas the residual cold gas (if any) is added to the ISM of the primary. 
The dark matter of the satellite halo is added to the dark matter of the primary halo. The halo gas of the satellite is added to the halo gas of the primary.}
We also tried the recipe in the SHARK semi-analytic model by \citet{Lagos2018} and found that 
the results are not significantly changed except for extremely low-mass halos 
for which the results are sensitive to the treatment of gas-rich mergers.

We performed calculation for 13 present halo masses equally spaced logarithmically between $10^9 \msun$   and $10^{15.5} \msun$  for the redshift range 0$<z<$9. 
For each halo mass, we performed 100 runs with different seeds for the random number generator to realize statistically different merger histories.
Each run thus has a unique history of mergers. In this paper, we exhibit the result of averaging those 100 runs after five iterations to examine the mean behaviors of the model for a given mass unless otherwise specified.
Five iterations were found to enough for sufficient convergence.
We leave  the study of statistical features brought about by mergers for future work.
We adopted WMAP1 cosmology \citep{Spergel2003} with the cosmological parameters $\Omega_{\rm M}=0.25$, $\Omega_\Lambda=0.75$, and $H_0=73 {\rm km s}^{-1} {\rm Mpc}^{-1}$.


\begin{figure*}
        \centering
        \includegraphics[width=1\linewidth]{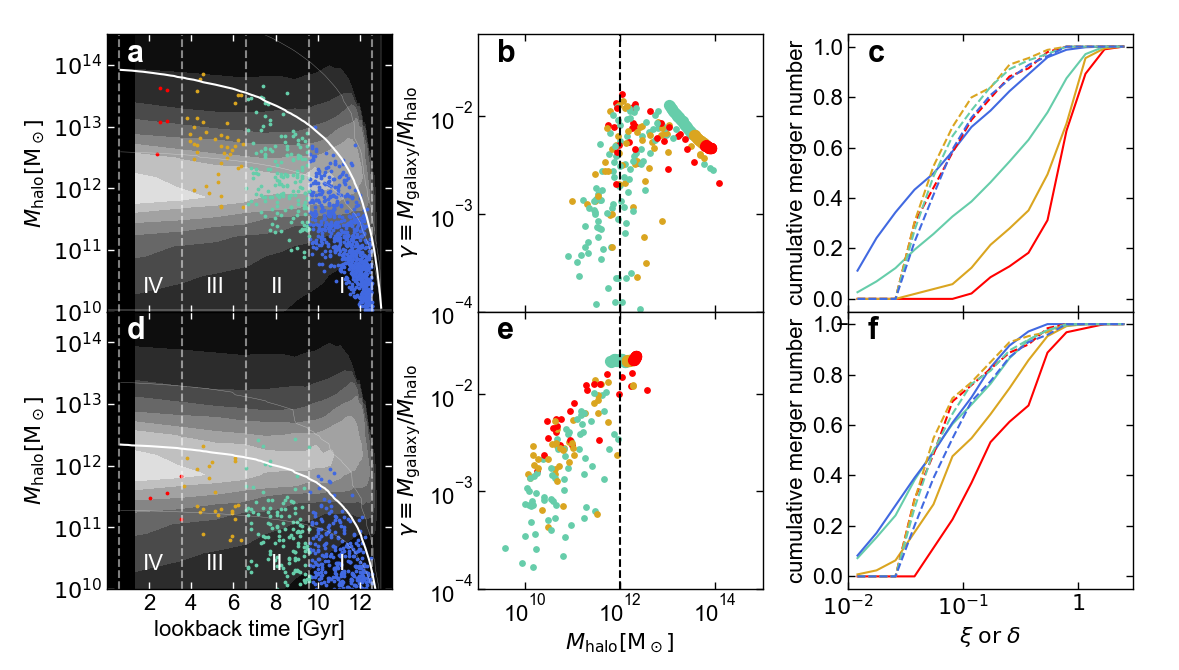}
        \caption{Merger characteristics for Galaxy A (upper panels) and Galaxy B (lower panels).
{\bf a} and {\bf d}: Masses of satellite halos are plotted at the time of mergers,
color-coded in accordance with epochs (I, II, III, IV) divided by vertical dashed lines. Solid lines indicate the mass growth of the primary halo (thick: mean, thin: maximum and minimum).
Superposed is the gray-scale map for the galaxy-to-halo mass ratio, $\gamma$, for
all galaxies in arbitrary units. {\bf b} and {\bf e}: Galaxy-to-halo mass ratios of the primary (large dots) and satellites (small dots) are plotted against halo masses, color-coded as in {\bf a} and {\bf d}.
The dashed vertical lines approximately indicate the peak mass in $\gamma-M_{\rm halo}$ relation.
{\bf c} and {\bf f}: Cumulative number distributions of $\xi$ (dashed lines) and $\delta$ (solid lines) for different epochs, color-coded as in {\bf a} and {\bf d}.}
        \label{fig2}
\end{figure*}

\maketitle
\section{The galaxy-to-halo mass ratios and the morphological sequence}
\label{feedback}

We have carried out four types of model calculation with or without feedback processes to examine their effects. 
Feedback processes put marked characteristics on the galaxy-to-halo mass ratios as already known and indicated below. Particular selection of feedback option is used as a tool for changing the resulting GHMR behavior
here and therefore we examine essentially the effect of GMHR on the galaxy morphology.
Model NULL includes no feedback processes. Model SN includes only SN feedback whereas Model AGN includes only feedback from AGN.
 Finally, Model SN+AGN includes both feedback processes. 
For Model SN and Model NULL, which include no AGN feedback, we set
$M_{\rm stream}$ large enough so that the hybrid accretion (i.e., half the halo gas 
accretes in free-fall and another half accretes in the radiative cooling time)
occurs in the whole domain with $M_{\rm halo} > M_{\rm shock}$.
Namely, the filamentary streams of unheated gas are preserved for this domain.

We first compare model results and observational data regarding the present-day morphological sequence.
Fig. 1 shows GHMR along with  the bulge-to-total stellar mass ratio ($B/T$) and ages. Four Models are shown by different colors.
It should be noted that the galaxy means the sum of stars and cold interstellar medium, and the halo includes dark matter and hot halo gas in addition to the galaxy. 
Figs. 1a and 1d show that the galaxy-to-halo mass ratio, $\gamma$, for observed galaxies has a peak at 
$M_{\rm halo} \sim 10^{12} \msun$ from $z=$2 to 0 (from $\sim 10$ Gyr ago to the present). 
It is seen that Model SN+AGN gives the best match with the observation. Models lacking SN feedback produce too massive galaxies in low-mass halos whereas Models without AGN overproduce stars in high-mass halos. 
Model NULL suffers from both defects.

Figs. 1b and 1c show that Model SN+AGN best reproduces the variation of the bulge fraction ($B/T$) with galaxy mass for local galaxies. 
Models without SN feedback produce too large $B/T$ for low-mass galaxies whereas models without AGN feedback fail in reproducing large $B/T$ in massive galaxies. 
We thus need both SN and AGN feedback to get the correct morphological sequence. 
This result itself is not surprising because the fundamental physics included in Model SN+AGN is designed essentially in the same manner as those employed in past numerical and semi-analytic studies mentioned above 
which reproduced the observed morphological sequence. 
The analysis here indicates that the specific form of 
$\gamma - M_{\rm halo}$ relation is the direct cause of the observed morphological sequence. 
In passing, we note that Model SN+AGN performs best also in reproduction of the age variation of bulges (Figs. 1e,1f). We revisit this point in section 5.

\begin{figure*}[htb]
        \centering
        \includegraphics[width=1\linewidth]{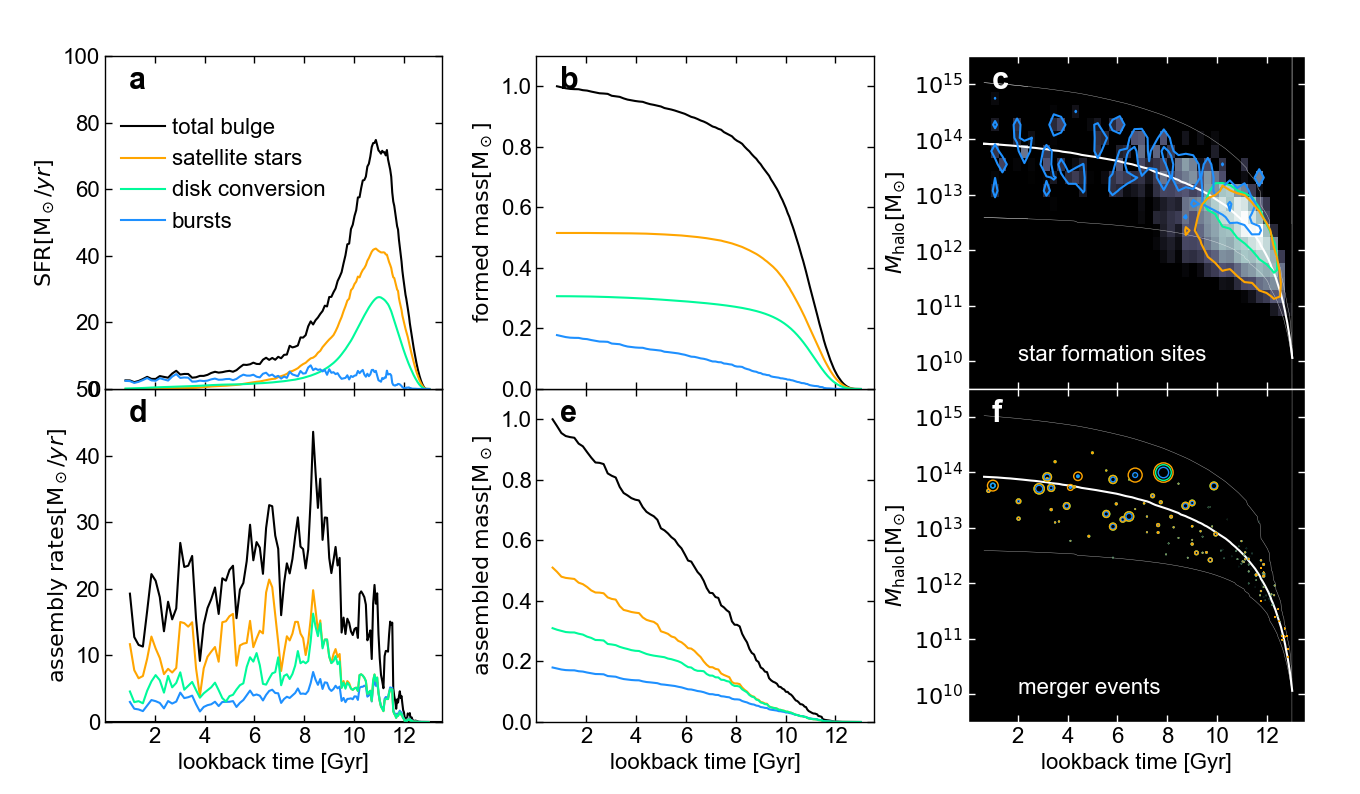}

        \caption{Star formation history and assembly history for bulge stars in Galaxy A.
Three pathways of the bulge growth are shown separately
(green: conversion of the primary disk, orange: acquisition from satellite galaxies, cyan: starbursts driven by mergers).
Total bulge component is shown in black. {\bf a} and {\bf d} show the star formation rates (SFR) and assembly rates, respectively.
{\bf b} and {\bf e} respectively plot the growth of the formed mass and the assembled mass. The gray-scale map in {\bf c} shows the formation sites of bulge stars, i.e., when and in what mass of the halo they were formed.
Colored contours indicate 20\% level of the peak intensity for each pathway. {\bf f} indicates assembly (i.e., merger) events.
Radii of the colored circles are proportional to the assembled mass for three pathways.
In {\bf c} and {\bf f}, white solid lines indicate the mass growth of the primary halo (thick: mean, thin: maximum and minimum).}
        \label{fig3}
\end{figure*}

\maketitle
\section{Merger demography and bulge build-up}
\label{demography}

We next explore how the peaked $\gamma - M_{\rm halo}$ relation affects bulge mass fractions in the presence of mergers and creates the observed morphological sequence. 
Variation of the bulge mass fraction, $B/T$, is essentially controlled by the mass ratio of the satellite galaxy 
and the primary galaxy, 
$\mu = M_{\rm galaxy,s} / M_{\rm galaxy,p}$, 
because all stars in the satellite galaxy are added to the bulge and the same mass of the primary disks is converted into the bulge (i.e., the disk mass decreases) in our recipe described above. 
This argument neglects possible effect of gas components in merging galaxies.
As we see later in Fig. 4, however, the contribution of merger-driven starbursts 
is less than $\sim 30$ percent for $M_{\rm star} > 10^{10} {\rm M}_\odot$, the mass range 
which we are interested in because of large variation in
the bulge fraction $B/T$.
The ratio $\mu$  is related to $\gamma$ as follows.
\begin{equation}
\begin{aligned}
 \mu &= (\gamma_{\rm s} M_{\rm halo,s}) /(\gamma_{\rm p} M_{\rm halo,p} ) \\
 &= (\gamma_{\rm s} /\gamma_{\rm p})(M_{\rm halo,s}/ M_{\rm halo,p} ) \\
 &= (\gamma_{\rm s} /\gamma_{\rm p})\xi 
\end{aligned}
\end{equation}
, where $\xi \equiv M_{\rm halo,s}/ M_{\rm halo,p}$.
Therefore, $\mu$ is proportional to $\delta \equiv \gamma_{\rm s} /\gamma_{\rm p}$ for a fixed value of $\xi$.
The quantity $\delta$  measures the relative enhancement or decrement of the 
satellite galaxy-to-halo mass ratio over the primary galaxy-to-halo mass ratio. 
As this equation indicates, the galaxy mass ratio $\mu$ depends also on the halo mass ratio in a merger. 
The parameter $\delta$ thus isolates the effect of the galaxy-to-halo mass ratios in two merging halos.  It is more relevant than 
the galaxy mass ratio $\mu$ when we discuss change of $B/T$
using $\gamma - M_{\rm halo}$ relation, because it measures the relative vertical position 
of two merging halos on this relation.
Hereafter, we call this quantity the {\it baryonic bias} when convenient.

Hereafter we concentrate on Model SN+AGN. Fig.2 presents behaviors of two mass groups. 
We use here the word 'mass group' because we present below the mean behavior 
of 100 runs with different random number seeds for each prescribed halo mass 
 (see section 2 for detail). A single mass group is called 'Galaxy'.
Galaxy A represents 100 massive halos with the present mass $M_{\rm halo} \sim 10^{14} \msun$ and attains final bulge
 fractions of $B/T \sim0.8$, while Galaxy B with the present mass $M_{\rm halo} \sim 10^{12} \msun$ barely develops bulges. 
Figs. 2a and 2d present cosmological epochs of mergers and masses of involved satellite halos by
colored dots.
White solid lines indicate the halo mass growth in Galaxies A and B.
Underlying gray maps show the galaxy-to-halo mass ratios, $\gamma$, for Model SN+AGN as a function of halo mass and time.
We clearly see the ridge in $\gamma$ corresponding to 
the peak mass $M_{\rm halo} \sim 10^{12} \msun$ , which stays nearly constant with time. 
This is a different form of representing the $\gamma - M_{\rm halo}$ relation plotted in Fig.1.
The existence and horizontality of the ridge is important in elucidating why and how the galaxy morphological sequence emerged in presence of mergers.
Due to this configuration of the ridge,
 characteristics of mergers differ depending on whether the primary halo is located on the high-mass side or low-mass side of this ridge.  

Neglecting epoch I in which bulges grow little, Galaxy A, which stays above the peak mass
for most time, has high probabilities to merge  with  satellite halos located on the high-mass side of the ridge. 
Those satellites  have $\delta>1$
 owing to the peaked nature of the $\gamma - M_{\rm halo}$ relation and therefore 
enhance the bulge mass fraction efficiently.
The situation is conversed for Galaxy B.
It mostly  merges with satellites on the low-mass side of the peak having $\delta<1$ so that 
the growth of the bulge fraction is suppressed.
The  merger rate and $\xi$-distribution in cosmological dark matter simulations 
do not depend much on halo masses \citep{Fakhouri2010,Zhang2023}, 
though the merger rate has a weak dependence  of the form
 $M_{\rm halo}^{0.13}$ \citep{Fakhouri2010}.
Namely, the merger history of halos does not strongly depend on the halo mass
 and the discussion here which focuses only on $\delta$ is validated
 (see the discussion about Figs. 2c and 2f below).

Figs. 2b and 2e visualize how the primary and satellites move with time along the $\gamma-M_{\rm halo}$ relation. 
In both Galaxies A and B, satellites in epoch II (green small dots) are mostly distributed in the downward slope toward smaller halo masses. 
Satellites in Galaxy B  mostly remain on this slope afterward.  
On the other hand, the primary in Galaxy B remains near the peak of the $\gamma-M_{\rm halo}$ relation, 
indicating that most mergers in Galaxy B have $\delta <1$.
Galaxy A traces different evolution.
As we move from  epoch II to  III to IV, increasingly larger parts of satellites climb up the slope and come to lie around the peak, 
while the primary constantly goes down the rightward slope. 
In epoch IV, most satellites have $\delta>1$ (red small dots). 

Figs. 2c and 2f plot the distributions of $\delta$ (solid lines) and
halo mass ratios $\xi$ (dashed lines).
  These diagrams highlight the different behaviors of the baryonic bias in Galaxies A and B seen in Figs. 2b and 2e.
It is evident that two Galaxies have quite similar distributions in the halo mass ratio. 
Moreover, those distributions hardly change with time, reflecting similarities in merger history for different halo masses \citep{Fakhouri2010,Dong2022,Zhang2023}.
In contrast, the baryonic bias parameter $\delta$ behaves quite differently in two Galaxies. 
The distribution in Galaxy A becomes strongly skewed toward larger values with time while that in Galaxy B does not change much and stays close to the distribution of $\xi$.
These analyses reveal the key role of the baryonic bias in controlling bulge mass fractions in  mergers. The weak halo mass dependence,  
 $M_{\rm halo}^{0.13}$,  mentioned above leads to the increase in halo merger rates 
of only factor $\sim 1.8$ over 2 dex of the halo mass and therefore cannot fully  explain 
the increase of $\sim 7$ in $B/T$ shown in Fig. 1b.

How the nonlinear stellar-to-halo mass relation affects the characteristics of galaxy mergers has been 
discussed partly in past studies.
\citet{maller2008} pointed out that non-linear mapping of halo mergers to galaxy mergers provides possible explanation of why Hubble type depends on galaxy mass.
\citet{oleary2021},  
 using the empirical galaxy evolution model EMERGE,
 argued that more massive galaxies experience major mergers more frequently as a 
consequence of the nonlinear relation. Both studies, however, did not address 
the consistency between the observed stellar-to-halo mass relation and the $B/T$-mass relation of
local galaxies.
\citet{correa2020} 
report that disk and elliptical galaxies follow different stellar-to-halo mass relations  in 
both observations and EAGLE cosmological simulations. 
Their argument is that the SHMR  
is controlled by galaxy morphology, which is in  turn determined by the history of  gas accretion and star formation 
 and the degree  of AGN feedback. 
Their view is rather opposite to ours, which argues that the SHMR (actually GHMR) controls galaxy morphology 
with  mergers acting as mediators.

\begin{figure}
        \centering
        \includegraphics[width=1\linewidth]{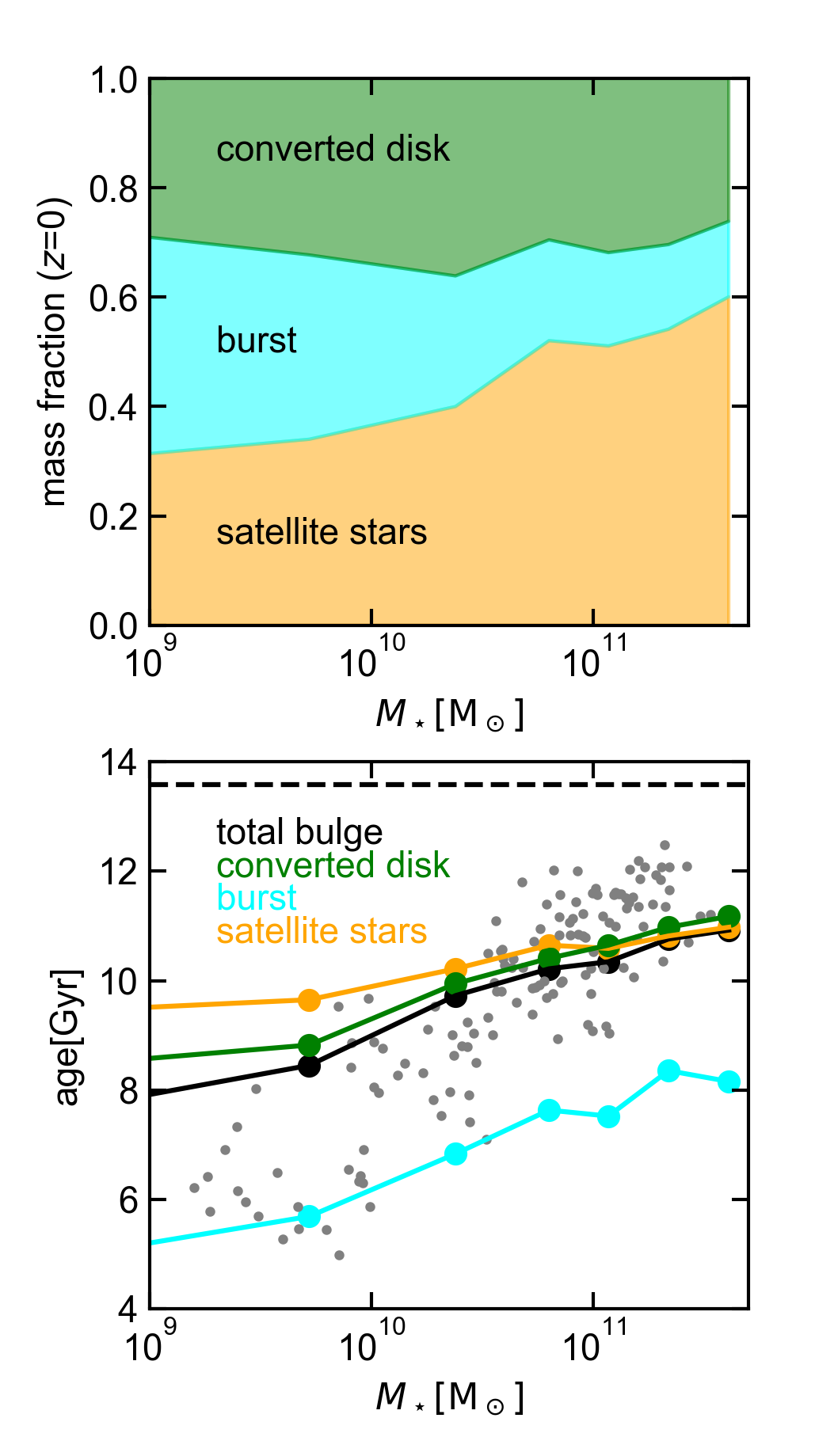}
        \caption{Mass fractions (upper panel) and ages (lower panel) of three components of the present bulge plotted against the present stellar mass of the galaxy. 
Gray dots in the lower panel indicate the observation for total bulges in local galaxies by \citet{Breda2018}.
              }
\end{figure}

\maketitle
\section{Formation and assembly of bulge stars}
\label{formation}

Having seen the general trend across galaxy mass, we now examine in more detail the process of bulge formation in Galaxy A. 
As already stated, bulges generally grow in three ways through mergers \citep[e.g.][]{delucia2007,Hopkins2009}. 
First is the direct acquisition of stars from satellite galaxies. 
Second, stars formed in merger-driven starbursts are added to  bulges.
Third is the conversion of the primary disks into bulges.
These processes are taken into account in many semi-analytical studies for galaxy formation \citep[e.g.][]{somerville1999,croton2006,Lagos2018}.
In analyzing the bulge formation process, it is important to discriminate between the formation time and assembly time of bulge stars, 
which are not necessarily the same \citep[e.g.][]{delucia2007}. 
Accretion of satellite stars and conversion of the primary disk are delayed 
with respect to the formation of involved stars.
These two times are identical for stars born in starbursts.

Fig.3 describes the star formation history and the assembly history of the bulge in Galaxy A.
Three pathways of bulge growth are shown by different colors as indicated.
Fig.3a, which plots the star formation rate, indicates that most bulge stars that originate in satellites and pre-existing primary disks were formed in early times. 
Mergers scatter and retard the assembly time of these stars as seen in Fig. 3d, which plots the assembly rates. 
Formation and assembly times of stars created by bursts are distributed over wide range of times 
following the broad time distribution of mergers.  Figs 3b and 3e show the cumulative distribution of formation rates (SFR) and assembly rates of bulge stars.
 In total, half the bulge stars were formed more than 10 Gyr ago but they joined the bulge 6 Gyr ago at last. 
In other words, the bulge in this Galaxy is old but appeared as an integrated structure relatively recently in the history of the universe. 
It is noted that, in Galaxy A, about half the bulge stars are acquired from satellites
and the contribution of merger-induced bursts is less than  20 percent.
These features for massive bulges, i.e., a large delay of assembly and unimportance of bursts, agree very well with the previous SAM result
obtained by \citet{delucia2007} for the brightest cluster galaxies.

It is enlightening to examine the formation sites and the assembly history of bulge stars.
Fig. 3c, which shows the formation sites of bulge stars in gray scale, demonstrates that 
most bulge stars were formed in progenitor halos with $M_{\rm halo} \sim 10^{12-13} \msun $ more than 10 Gyr ago. 
As shown in Fig. 3f, which plots the merger events, assemblage of bulge stars 
is delayed with respect to their formation 
 and proceeds over a long period while the primary halo grows up to $\sim 10^{14} \msun$.

Fig. 4a shows the contribution of three channels of bulge formation to the present-day bulge masses as a function of the galaxy stellar mass at present.
We see a clear trend that the direct addition of satellite stars 
which is the primary growth channel  for massive galaxies is gradually taken over 
by starbursts driven by mergers as the galaxy mass decreases. 
This is naturally expected because  low-mass galaxies grow by mergers with galaxies having even lower masses, which tend to be richer in gas than the satellite galaxies acquired by massive galaxies. 
Also the low-mass primary itself tends to be richer in gas than the high-mass primary.
The contribution of disk conversion stays nearly constant with the galaxy mass.
As seen from this figure, the contribution of merger-driven starbursts remains 
subdominant expect for low-mass galaxies with $M_{\star} < 10^{10} {\rm M}_\odot$, 
which are disk-dominated.

\begin{figure*}[htbp]
        \centering
        \includegraphics[width=1\linewidth]{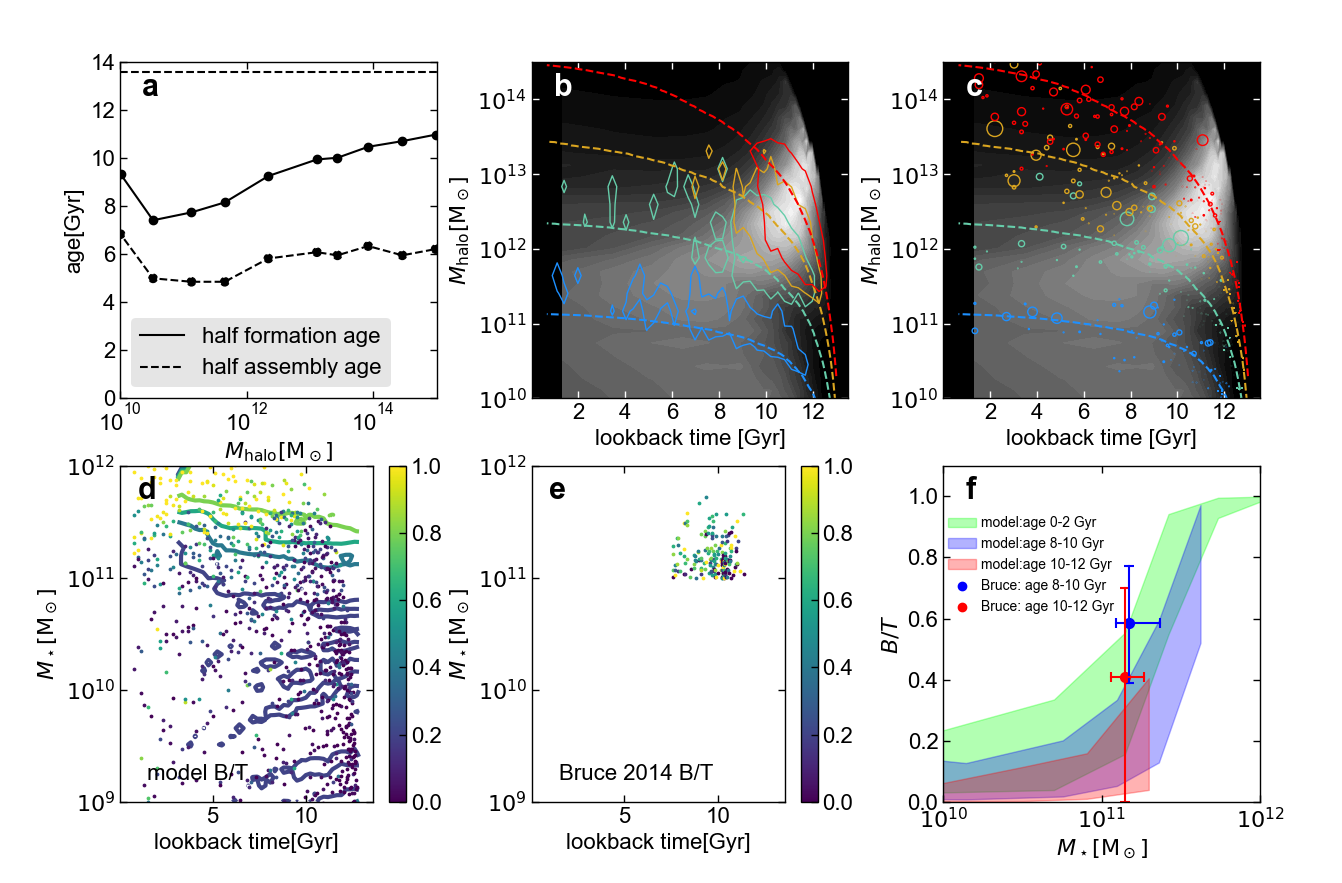}
        \caption{Upper panels: Halo-mass dependence of bulge formation and assembly. {\bf a}: Solid (dashed) line indicates the formation (assembly) age, which is the lookback time at which half of the stars contained in the present bulge were formed (assembled).
{\bf b}: Contours indicate the formation sites of the present bulge stars for four mass groups by different colors. Each contour shows the level for half the maximum value.
Dashed curves indicate the mean growth of halo mass in each group. The gray-scale map indicates the  star formation rate as a function of the halo mass and the lookback time for all galaxies  in arbitrary units.
{\bf c}: Same as {\bf b} but merger events are indicated by circles, the radius of which is proportional to the ratio of the satellite galaxy mass and the present bulge mass.
Lower panels:
{\bf d} and {\bf e}: $B/T$ ratios for model galaxies ({\bf d}) and CANDELS galaxies observed by Hubble Space Telescope from \citet{Bruce2014} ({\bf e}).
In {\bf d}, the contours for $B/T$ = 0.2, 0.4, 0.6 0.8 are drawn with corresponding colors.
{\bf f}: Development of the galaxy morphological sequence.
 Shaded regions indicate  25 and 75 percentiles for model galaxies
at each galaxy stellar mass for different lookback times as given in the diagram.
Dots indicate the median in the stellar mass and $B/T$ for CANDELS galaxies while horizontal and vertical bars indicate 25 and 75 percentiles for the galaxy mass and $B/T$, respectively,
for two indicated epochs.
       }
        \label{fig5}
\end{figure*}

Alongside the morphological sequence, another important observational result is the age sequence of bulges: Bulges of more massive galaxies are older \citep[e.g.][]{Breda2018,Parikh2024}. 
Fig. 4b (also see Figs. 1e and 1f) indicates the ages of total bulges and three  components of different origins.
Bulges in total show the mass dependence similar to the observed one though the latter is steeper.
Satellite stars constitute the oldest part of bulge stars irrespective of galaxy mass while 
burst stars are younger by $\sim 3 $Gyr than the total bulges. In low-mass galaxies, significant contribution  of burst components 
makes the total bulge age younger than those in high-mass galaxies.
For low-mass galaxies, the observed bulges are younger than the model total bulges by a few Gyr.
Possible contribution from pseudo bulges is likely to make this discrepancy as we discuss in the end of this section.
It should be noted that the age estimate for old stellar populations is generally difficult. 
Moreover, difficulty in observing neutral hydrogen gas in distant galaxies hampers reliable estimate of cold gas content in early times, causing ambiguity in model prediction
for early star formation rates.
Upcoming Square Kilometre Array project \citep{Staveley-Smith2015} is expected to improve this situation and enable more reliable comparison.

Fig. 5 summarize the formation and assembly history of bulge stars for different halo masses to aid the interpretation of the age sequence.
Fig. 5a compares bulge formation ages and assembly ages. It is seen that the latter 
is smaller than the former by several Gyr and does not show strong mass dependence
in contrast to the formation age.
As seen in Fig. 5b, which shows the formation site of all bulge stars by colored contours, halos with lower present masses possess bulge stars formed in progenitor halos with lower masses in more recent times. 
Note that the superposed gray-scale map indicates the star formation rate distribution
for all the model galaxies.
Fig. 5c, which overplots merger events on the SFR map and illustrates bulge assembly process, explains the origin of bulge age sequence obtained here. Massive galaxies gather satellite galaxies that made most stars in well past. 
On the contrary, low-mass galaxies gather satellite galaxies, the star formation activity of which is still rising at the time of mergers, leading to their young bulges. 
Fig. 5c indicates that the assembly events are widely scattered over time regardless of halo mass. 
This leads to assembly ages several Gyr younger than formation ages as shown in Fig. 5a. 
Weak dependence on the halo mass reflects the similarity in merger history. 
Several recent studies discuss formation history and the age sequence of galactic bulges in some detail \citep[e.g.][]{Breda2018,breda2020,constantine2021}
These studies examine total bulges. 
The analysis performed here based on Figs. 4 and 5 deepens our understanding of 
the bulge age  sequence by dividing constituent stars into three components with different origins.

Fig. 5d plots the bulge-to-total mass ratio $B/T$ for individual model galaxies together with contours for constant $B/T$.
Observational data to compare with the model prediction is difficult to get currently due to small apparent sizes and 
low surface brightness of distant galaxies \citep{Brennan2015,Lang2014,Buitrago2013}. 
Figs. 5e and 5f plot the sample of galaxies  observed by HST and analyzed by \citet{Bruce2014}. 
The blue and red dots in Fig. 5f indicate the median  $B/T$ derived from data for individual galaxies plotted in Fig. 5e. 
Vertical and horizontal bars indicate 25 and 75 percentiles for $B/T$ and the stellar mass, respectively.
Color-shaded areas show 25 and 75  percentiles for $B/T$ at each stellar mass derived from individual model
galaxies in Fig. 5d.
Figs. 5e and 5f indicate that observed massive galaxies ($M_\star \gtrsim 10^{11} \msun$) around 10 Gyr ago already have diverse morphology 
with mean $B/T \sim 0.5$. 
The model agrees qualitatively with the observation but seems to show somewhat smaller $B/T$ at two common epochs. 
However, the model $B/T$ increases sharply at this mass scale so that more observation for lower and higher masses is required for critical comparison.
Recently launched JWST \citep{Gardner2023}, with ability to observe galactic structures for a wider domain of mass and redshift than HST, is expected to reveal how the galaxy morphological sequence developed with time in greater detail.

Finally we make one remark on the comparison of model and observational ages performed here.
Bulges younger than $\sim 8$ Gyr in local small-mass galaxies in the Breda-Papaderos sample plotted in Fig. 4b 
are significantly displaced from the model prediction.
In this study, we have concentrated on the formation of classical bulges which are 
considered to be products of mergers. 
Another category of galactic bulges is pseudo-bulges with flatter shapes and younger ages
\citep[e.g][]{gadotti2009}. It has been proposed that these are formed through secular evolutionary 
processes such as action of bar structures \citep[e.g.][]{kormendy2004} and viscous mass transfer by
massive gas clumps in gas-rich disks \citep[e.g.][]{noguchi1998,noguchi1999}.
 Pseudo-bulges tend to reside in low-mass galaxies.
The bulge demography by \citet{fisher2011} indicates that pseudo-bulges take over 
classical bulges at ${\rm log} (M_{\rm star}) \sim 10.5$.
Young bulges in Breda-Papaderos sample 
  may thus be pseudo-bulges.

\section{Conclusions}
Using a simple analytical model of galaxy evolution,  we have explored the effect of the halo-mass dependence of 
the galaxy-to-halo mass ratio on the galaxy morphology in presence of mergers.
This ratio  has a peak   located at $M_{\rm halo} \sim 10^{12} \msun$ almost independent of the redshift. 
Massive halos spend long time above this peak mass and therefore preferentially merge with 
satellite halos with larger galaxy-to-halo mass ratios than themselves. 
Those satellites increase the bulge-to-total mass ratios efficiently.
Bulge growth is limited in 
low-mass halos, which evolve mainly below the peak mass and merge with satellites 
having relatively smaller galaxy-to-halo mass ratios.
This means that the peaked shape of the galaxy-to-halo mass ratio plays an important role in 
establishing the morphological sequence of galaxies.
Because the peaked nature  of the galaxy-to-halo mass ratio is most likely created by 
feedback processes from young massive stars including supernovae and active galactic nuclei,
the present result suggests that those feedback processes are closely linked 
to the galaxy morphological sequence.
The predicted build-up history of bulge components is qualitatively consistent with  
currently available limited data but observations for more extended time and galaxy (or halo) mass are required for critical tests of the model.

\section*{Data Availability}
 The data that support the findings of this study are available from the corresponding author upon reasonable request.

\section*{Code availability} 
We have opted not to make available the code used to calculate the evolution of galaxy models because it is a part of the integrated program currently in use for other projects.

\section*{Acknowledgments}
 MN acknowledges Shy Genel and anonymous referees for helpful comments.

\bibliography{main}{}
\bibliographystyle{aasjournal}


\end{document}